\documentclass[%
 ,twocolumn%
 ,secnumarabic%
,amssymb,amsmath,aps,prb,floatfix,showpacs]{revtex4-1}
\usepackage{docs}%
\usepackage{bm}%
\usepackage[colorlinks=true,linkcolor=blue]{hyperref}%
\usepackage{latexsym}
\usepackage{amsmath,amssymb,amsfonts,textcomp}
\usepackage{graphicx}
\usepackage{natbib}
\usepackage{mathrsfs} 
\usepackage[T1]{fontenc}

\def\b#1{\mathbf{#1}}

\def\m#1{\mathrm{#1}}
\def\ss#1{\scriptscriptstyle{#1}}

\def\u0#1#2{u^{\scriptscriptstyle{\mathrm{\;#1}}}_{\scriptscriptstyle{0#2}}}

\begin{document}

\title{The low temperature elastic anomalies in solid helium}%
\author{Eric Varoquaux}%
\email{eric.varoquaux@cea.fr}
\affiliation{CNRS and CEA-IRAMIS-DRECAM, Service de Physique de l'\'Etat Condens\'e, \\
  Centre d'\'Etudes de Saclay, 91191 Gif-sur-Yvette Cedex (France)}
\date{\today}%

\begin{abstract}
  The elastic properties of hcp $^4$He samples have been shown to display
  various anomalies. As the temperature is lowered below $\sim$\,0.2 K, the
  elastic shear modulus appears to stiffen and the moment of inertia appears
  to drop in a concomitant manner. The former phenomenon is taken as evidence
  for the pinning of dislocations, the latter, for the appearance of
  supersolidity. The close relationship between these two observables is
  studied within the framework of classical deformable-body mechanics. A model
  based on the formation by plastic flow of extremely soft quasi-planar,
  inter-connected layers of dislocations is solved analytically and
  numerically. This model relates quantitatively the change in moment of
  inertia to the drop in elastic constant and can account for most
  experimental observations. Other situations, in which its relevance may seem
  more doubtful, are discussed.
\end{abstract}
\maketitle
\clearpage

\section{Introduction} \label{Introduction} 
Helium makes an intriguing solid. Both $^4$He and $^3$He isotopes crystallise
from the liquid at absolute zero under a pressure of 24.5 and 34.5 bars
respectively. They form hexagonal close-packed (hcp)
crystals with similar elastic properties. These crystals are very soft, owing
to their small densities and weak atomic interactions. The paucity of slip
directions causes crystals with the hcp structure to be quite brittle: they
fracture very readily.  This combination of softness and brittleness in the
helium crystals make them very prone to plastic deformation and the formation
of dislocation lines.

Because of the fast boson exchange in solid $^4$He and the presence of
defects, the possibility that a Bose-Einstein condensate would form within the
crystalline lattice below a certain temperature was raised by a number of
authors, starting with Penrose and Onsager in the fifties.\cite{Penrose:56}
These authors argued that superfluid coherence, or off-diagonal long-range
order (ODLRO), would not occur in an ideally perfect crystal but possibly
could in a distorted lattice. Although the proof that they gave
was criticised by others,\cite{Andreev:69,Chester:70,Leggett:70} it marked the
beginning of a long-lasting search, both theoretical and experimental, for
features that could reveal the existence of such a ``supersolid'' state in a
suitably disordered crystal.

This search received strong renewed impetus after the observation by Kim and
Chan (see the reviews [\onlinecite{reviews}]) of an anomaly in the rotational
inertia of $^4$He solid samples as seen as a period shift in
high-quality-factor torsional oscillators (TO). The increase in the period of
the oscillator resonance below a temperature of $\sim$ 0.2 K, now observed by
many groups, is taken to signal the decoupling of part of the helium mass from
the motion of the oscillator bob. This effect, first discussed by
Leggett\cite{Leggett:70} and called in the present context non-classical
rotational inertia (NCRI), occurs in a number of TO experiments with widely
different sizes and geometries including very confined ones such as
those of Vycor or sintered gold.

In the framework of the time-honoured two-fluid model for
superfluidity,\cite{Landau:63} such an observation would signal the appearance
of a superfluid-like fraction in the solid.  Such a ``condensate'' would
settle to rest and decouple from the oscillator walls as the temperature is
lowered, reducing the moment of inertia. This interpretation is born out by
the fact that, if the oscillator geometry is modified by a partition blocking
the closed loop along which the superflow is supposed to take
place,\cite{Kim:04,Rittner:08} the effect disappears. Also, NCRI is not observed
when the oscillator is filled with $^3$He instead of
$^4$He,\cite{Kim:04,West:09} which constitutes a strong hint that quantum
statistics plays a fundamental role.

The TO measurements do seem to suggest that some form of superfluid behaviour
occurs in solid $^4$He below 0.1$\sim$0.2 K but other unambiguous
manifestations of the existence of a true superfluid component, such as a
non-dissipative DC-flow,\cite{Sasaki:07a,Day:07b,Rittner:09,Ray:09} or a
persistent current, a second sound\cite{Kwon:09} or fourth
sound\cite{Aoki:08b} mode, the fountain effect, the signature of a BEC
condensate on neutron\cite{Wallacher:05,Bossy:10,Diallo:11} or X-ray
diffractograms\cite{Burns:08} are still lacking in spite of the efforts and
ingenuity of many research groups.

Shear modulus measurements in solid helium at low temperature provide another
class of anomalous elastic properties. These measurements span many years,
starting with the early work of \citet{Wanner:76} soon followed by others.
\cite{Tsymbalenko:78,Tsymbalenko:79,Iwasa:79,Iwasa:80,Paalanen:81} They have
recently been extended to the same range of temperatures and $^3$He impurity
concentrations as the TO experiments.
\cite{Day:07,Day:07c,Mukharsky:09,Rojas:10} A marked decrease in the shear
modulus $G$ takes place in most samples of hcp $^4$He upon warming from
$T\sim0$. The magnitude of the softening varies from sample to sample,
depending in particular on the $^3$He impurity content $x_3$ and the cooling
history. The drop in $G$ can be spectacular, down to less than 20 \%
from the $T\simeq 0$ value.\cite{Tsymbalenko:78,Paalanen:81,Mukharsky:09,Rojas:10} 

Day and Beamish\cite{Day:07} have argued that the $T$ and $x_3$ dependence of
$G$ were mimicking closely those of the period shifts in TO experiments. In
fact, the striking similarities between the shear modulus and the TO resonance
frequency drops make it hard not to believe that the two phenomena are somehow
related. Direct experimental studies of this possible connection have led to
diverging conclusions.\cite{Reppy:10,Kim:11,Mi:11,Reppy-private-communication}

This article \cite{PostonCondMat} outlines one possible such link between
those two different mechanical properties of hcp $^4$He. It differs from
similar attempts by other workers \cite{Nussinov:07,Yoo:09,Iwasa:10} because
it recognises from the start that the large drop in $G$ requires the bunching
of dislocation lines into extended quasi-planar highly deformable sheets, as
described in Sec.\,(\ref{Model}). The consequences of these assumed defects
are derived analytically for the shear modulus drop and for the apparent
change of inertia in Sec.\,(\ref{Results}). These two quantities can thus be
directly linked to one another. Numerical values are derived in
Sec.\,(\ref{Discussion}), where it is shown that this simple model may account
quantitatively for a number of experimental observations. The model does not
explain readily certain classes of experiments, notably those in confined
geometries, and also the absence of inertia anomaly in solid $^3$He. Some
speculations on these topics are offered in Sec.\,({\ref{Conclusion}).

\section{Soft layer model} \label{Model} 
\subsection{Planar layers of dislocations}
As mentioned above, the two helium crystals are very soft: the
longitudinal and transverse sound velocities, in the 200 to 500 m/s range, are
low. They are also very fragile: the yield strength is of the order of 0.2$\sim$0.5
bar,\cite{Suzuki:73,Suzuki:77} or very much less depending on the experimental
conditions.\cite{Sanders:77} Dislocations appear readily under very weak
mechanical perturbations or thermally induced stress.  Plastic flow takes
place during the formation and subsequent cooling of the solid helium
sample. In the process, dislocations form and migrate.

Early sound propagation measurements in hcp $^4$He in the 5-50 MHz frequency range
\cite{Wanner:76,Hiki:77,Iwasa:79,Tsuruoka:79} have revealed an anomalous
temperature dependence below 1 K of the longitudinal sound velocity.
This anomaly has been attributed to the unpinning of the dislocation
lines as the trapped $^3$He impurities escape from the dislocation cores by
thermal activation. This interpretation is well documented through the work of many
authors \cite{Tsymbalenko:79,Iwasa:80,Paalanen:81} and, more recently, by
\citet{Syshchenko:10}.

The analysis of the high-frequency sound propagation measurements 
yields typical values\cite{Iwasa:79} $\sim 10^6$ cm$^{-2}$ for the density of
dislocation lines $\Lambda$ and $5 \cdot 10^{-4}$ cm for the average distance
between the nodes of the dislocation network $L_{\m N}$, assumed random and
homogeneously distributed. The dimensionless quantity $\Lambda L_{\m N}^2$ is thus
found of the order of $\sim$\,0.25.

Lower frequency measurements\cite{Tsymbalenko:78,Tsymbalenko:79,Paalanen:81}
interpreted in the same manner with the help of the following
relation for the change of the effective shear elastic constant $G_{\m{eff}}$
relative to the $T=0$ value $G$ ,\cite{Granato:56}
\begin{equation} \label{ShearModulusGranatoLucke}
  G/G_{\m{eff}} = 1+ 24(1-\nu)\Omega\Lambda L_{\m N}^2/\pi^3  \; ,
\end{equation} 
give, assuming a value of 0.3 for the Poisson ratio $\nu$ and with the highest
value for the orientation parameter $\Omega \leq 1/2$, much larger values of
the quantity $\Lambda L_{\m N}^2$.  The shear modulus measurements by
\citet{Paalanen:81} were carried out at a low frequency of 331 Hz
and lead to a value of $\Lambda L_{\m N}^2$ $\gtrsim$ 1.0 to 2.5 depending on
samples. More recent measurements \cite{Day:07,Mukharsky:09,Day:09,Day:10}
have confirmed these results.  A softening of 86 \% has been observed in an
ultra-pure monocrystalline $^4$He sample by \citet{Rojas:10} at
frequencies in the 10-20 kHz range. In this extreme situation, the quantity
$\Lambda L _{\m N}^2$ would exceed 20 using the same values for $\nu$ and
$\Omega$ as above.

These values of $\Lambda L_{\m N}^2$, obtained at long wavelengths, are much
larger than the upper limit for a dense hexagonal network of dislocations,
which espouses the underlying lattice symmetry. As shown in the Appendix, this
limit is $1/\sqrt{2}$ for an ideal hcp network. The corresponding upper limit
of $G/G_{\m{eff}}$ as given by Eq.(\ref{ShearModulusGranatoLucke}) is 1.2,
which falls short of
observations\cite{Paalanen:81,Tsymbalenko:78,Mukharsky:09,Rojas:10} by a wide
margin: edge dislocation lines escape from their preferential homogeneous
hexagonal network structure and become quite extended.

This anomaly clearly points towards the formation of inhomogeneous dislocation
structures. The dislocation lines collect in dense arrays, such as the mosaic
structure that form along the boundaries between grains with slightly
misaligned lattice vectors,\cite{Friedel:67} or, more generally, in extended
planar structures. This rearrangement takes place during the formation of the
hcp $^4$He samples and under thermal stress during cool-down.

It has been shown by numerical simulations of dislocation dynamics, notably by
Amadeo and Ghoniem,\cite{Amodeo:90,Zbib:98} that dislocations collect into
different planar structures according to different applied
perturbations. Planar arrays composed of sets of dislocation dipoles lying in
planes containing the direction of the critical resolved shear stress form
under monotonic stress conditions. Other types of structures, slip bands of
parallel dislocation lines or dislocation cells, may appear under cyclic
perturbations, provided, {\it{e.g.}}, by mechanical vibrations.  These planar
defects have been observed in a number of metallurgical
samples.\cite{Friedel:67} Their phenomenology is well documented, as
reviewed, {\it{e.g.}}, by \citet{Takeuchi:76} and others. Such dislocation
substructures have also been observed in hcp $^4$He by X-ray topography by
\citet{Iwasa:95} and by transmission Laue diffraction by \citet{Bossy:10a}.

These defect structures are thicker than the Franck networks that separate two
grain boundaries of low-tilt angle. They are quite different from the random
network assumed in the Granato-L\"ucke model,\cite{Granato:56} as already
mentioned. They can be viewed as resulting from the propagation of dislocation
pileups under thermal stress in a way similar to the formation of cracks in
usual hcp metals.\cite{HirthLothe} Solid helium exists only under positive
pressure and does not actually crack. Other types of extended defects
appear and enable the crystal to yield in the deformation directions imposed
by the rigid wall boundaries.

The dislocation arrays formed in such a manner are densely packed and have a
high density of long dislocation lines; dislocations interact strongly and are
organised in extended structures of parallel lines. They become extremely
mobile when unadorned of the $^3$He impurities that pin them to the lattice at
$T\lesssim 0.2$ K and when unhampered by thermally excited phonons that
prevail at $T\gtrsim 0.8 $ K. Following the same line of reasoning that
leads to Eq.(\ref{ShearModulusGranatoLucke}), the resulting large values of
$\Lambda L^2_{\m N}$ lead to very soft and easily deformable layers. These layers
separate regions with depleted dislocation densities but of enhanced
crystalline quality in which deformation also occurs quite
readily,\cite{Rojas:10} at least in the directions of easy glide, but with the
geometrical limit $\Lambda L^2_{\m N} \lesssim 2^{-1/2}$.

\subsection{Strain standing waves: homogeneous case}
\label{HomogeneousCase}
The simple model to be studied below assumes the existence of quasi-planar
dislocation structures that facilitate both plastic and elastic deformations.
To make the problem easily tractable analytically, a fully-planar geometry is
assumed: the helium sample is taken to be confined between two parallel plane
walls located at $z=0$ and $z=R$ and extending to infinity along the $x$ and
$y$ axes. The deformation $u$ induced in the sample depends on $z$ and $t$
only (see Fig.\,\ref{waveform}); the problem is one-dimensional and easily
solvable.

Shear stresses and strains are produced in the sample either by moving one
plane, {\it e.g.}, that at $z=0$ (which would be the transmitter in the shear
modulus experiment) with respect to the $z=R$ plane, held steady (which
would be the receiver). Torsional oscillator experiments are mimicked by
moving both bounding walls in unison, letting the sample inertia develop
internal stresses.

If the helium sample is homogeneous with a density $\rho$ and a shear modulus
$G$ independent of position and time (no visco-elastic effect, no internal
structure), the deformation $u(z,t)$ obeys the following partial differential
equation:
\begin{equation} \label{SimpleWaveEquation}
  \rho \frac{\partial^2{u}}{\partial t^2} = G \frac{\partial^2{u}}{\partial
    z^2} \; .
\end{equation} 
This equation describes the propagation of transverse plane waves with
dispersion relation $\omega^2 = c_{\m T}^2 k^2$ and $c_{\m T}^2 =
G/\rho$. The harmonic solution of Eq.(\ref{SimpleWaveEquation}) at frequency
$\omega/2\pi$ is the sum of two counter-propagating waves:
\begin{equation} \label{SimpleWaveEquationSolution}
   u(z,t) = \left( \u0{}+ \m e^{-ikz} +  
             u_{\ss 0-} \m e^{ikz}\right) \m e^{i\omega t} 
          = u(z) \m e^{i\omega t}  \; .
\end{equation} 
The constants of integration for shear measurements  $\u0S+$ and 
$\u0S-$ are then given by
\begin{equation}         \label{SimpleConstantsShear}
  \frac{\u0S+}{\m  e^{ikR}} = \frac{-\u0S-}{\m  e^{-ikR}} 
    = \frac{\u0{ }{ }}{ \m e^{ikR} -  \m e^{-ikR}} = \overline{\u0{}{}}
  \; ,
\end{equation}
and the solution of Eq.(\ref{SimpleWaveEquation}) for the deformation
as a function of $z$ can then be expressed under the following form:
\begin{equation}         \label{SimpleSolution}
  u(z)       = \u0{}{} \frac{\sin k(R-z)}{\sin kR} \; .
\end{equation}
The stress in the solid is derived from the deformation, still
disregarding the time dependence:
\begin{equation}         \label{SimpleStress}
  \sigma(z) = G\; \frac{\m d u}{\m d z}
                  = -k \u0{}{} G \frac{\textstyle\cos k(R-z)}{\textstyle\sin kR} \; .
\end{equation}
The force per unit area acting on the receiver is the opposite of that acting
on the body, namely the internal stress:
\begin{equation}         \label{ReceiverForce}
  F_{\m R} = -  \sigma(R) =\frac{k\u0{}{}\;G}{\textstyle\sin kR} \; ,
\end{equation}
so that the measured effective shear elastic modulus $G_{\m{eff}}$ is such that:
\begin{equation} \label{SimpleEffectiveModulus} 
  \frac{G}{G_\m{eff}} =
    \frac{\u0{}{}}{R}\,\frac{G}{F_{\m R}}
    =\frac{\textstyle\sin kR}{kR} 
    \simeq 1 - \frac{\rho \omega^2 R^2}{6\; G}+\cdots\; .
\end{equation}
Equation (\ref{SimpleEffectiveModulus}) describes the change of the effective
shear modulus at finite frequency due to the elastic response of the body. In
the limit $\omega\longrightarrow 0$, $G_\m{eff}$ reduces to $G$. For $kR =
\pi$, the body is set into resonance and, as damping has been neglected, the
effective shear modulus diverges. Higher frequency modes are not considered here.

For torsional oscillator measurements this elastic response of the body
becomes the dominant effect. In these experiments the two walls at
$z=0$ and $z=R$ are set into identical motion $\u0{}{} \m e^{i\omega
  t}$. The solution to Eq.(\ref{SimpleWaveEquation}) that satisfies the
boundary conditions 
\begin{equation}        \label{InertiaBoundaryConditions}
  u(z)|_{z=0} = u(z)|_{z=R} = \u0{}{} 
\end{equation}
can be written with the help of the following relations
\begin{equation}         \label{SimpleConstantsInertia}
  \frac{\u0M+}{1-\m  e^{ikR}} =  \frac{-\u0M-}{1-\m  e^{-ikR}} 
    = \frac{\u0{}{}}{ \m e^{ikR} -  \m e^{-ikR}} = \overline{\u0{}{}}
  \; .
\end{equation}
In particular, the stress $\sigma(z)$ is found to be:
\begin{equation}         \label{SimpleStressInertia}
  \sigma(z) = G \frac{\m d u}{\m d z} 
         = -kG\u0{}{} \frac{\cos k(R-z) - \cos kz}{\sin kR} 
  \; .
\end{equation}
The quantity actually measured in the TO type of experiments is the
back-action of the sample on the measuring device, namely the total force
$F_{\m X}+F_{\m R}$ exerted by the solid helium on both walls. This force is
expressed, per unit area, by
\begin{eqnarray}         \label{SimpleInertia}
  F_{\m X}+F_{\m R} &=&  \sigma(0) - \sigma(R)  
        =  2kG\u0{}{} \frac{1 - \cos kR}{\sin kR} \\
       & \simeq & \rho R \omega^2 \u0{}{}\left[ 1 + \frac{\rho\omega^2R^2}{12 G} +
           \cdots\right] \; . 
   \label{SimpleInertiaExpanded}
\end{eqnarray}   
The meaning of Eqs.(\ref{SimpleInertia}) and (\ref{SimpleInertiaExpanded}) is
made clear by the prefactor of the right-hand side
Eq.(\ref{SimpleInertiaExpanded}): $\omega^2 \u0{}{}$ is the acceleration
amplitude, $\rho R$ the ``bare'' inertial mass $M_{\m I}$ per unit area and,
in the square bracket, the elastic correction at finite frequency. This
``effective mass'' correction increases with frequency up to the resonance at
$kR = \pi$ where it becomes very large.

The shear modulus and effective mass corrections are related through
Eqs.(\ref{SimpleEffectiveModulus}) and (\ref{SimpleInertia}). Taking, {\it
  e.g.}, $R=1$ cm, a frequency $\omega/2\pi=1$~kHz, at a density $\rho =
0.194$ g/cm$^2$ for which $c_{\m T}= 267$ m/s, the effect of shear elasticity
on the effective mass amounts to 4.6 10$^{-3}$, which is not insignificant. To
a drop by 20\% in $G$ corresponds an apparent change in the mass by $\sim
\;10^{-3}$.

\begin{figure}[t]
  \includegraphics[width=80mm]{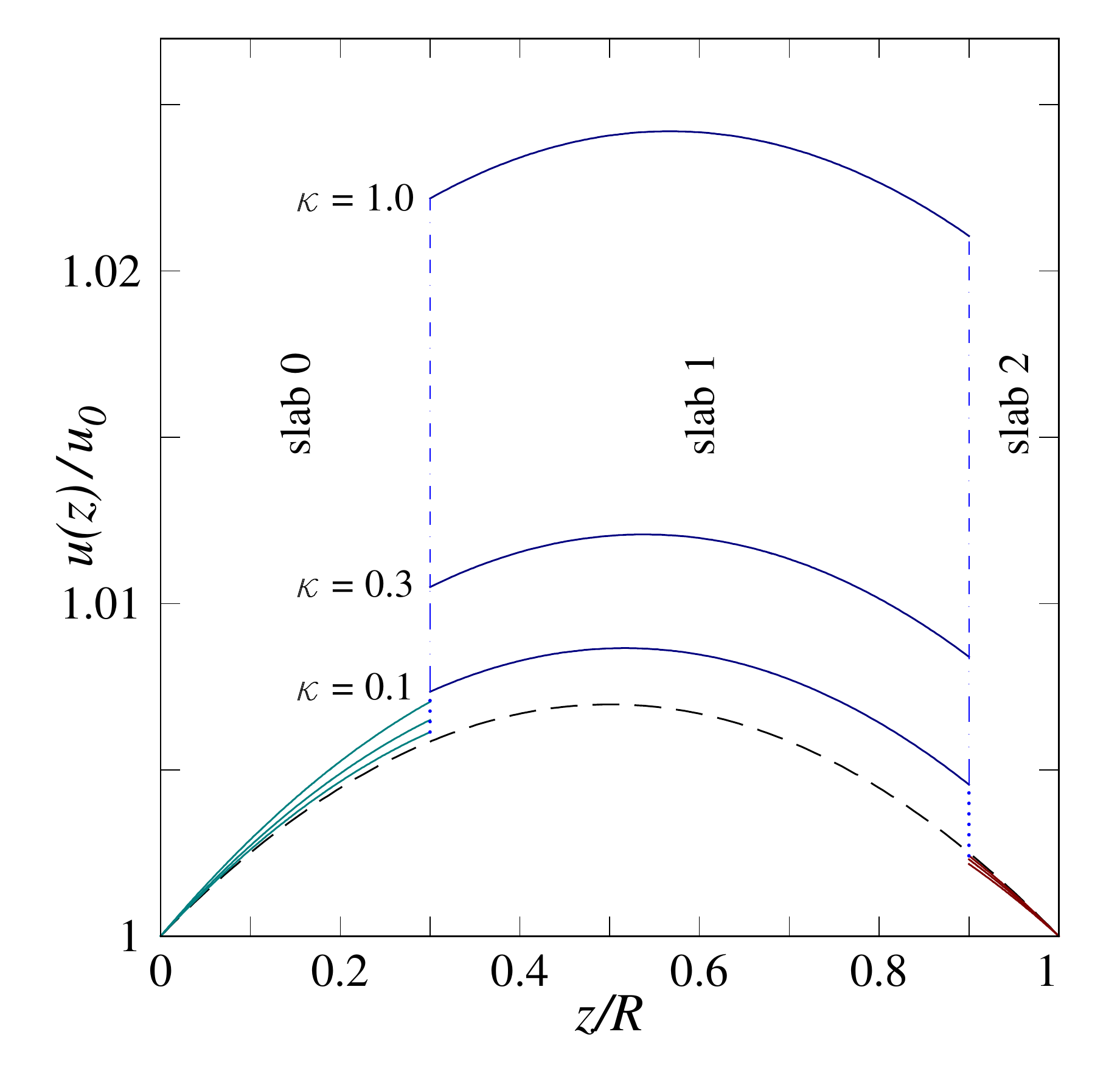}
  \caption{\label{waveform}Forms of the stationary wave in TO's experiments
    pictured as the relative displacement $u(z)/u_0$ for $z$ varying from 0 to
    $R$ for various dimensionless compliances $\kappa$ of the soft layers,
    taken to be equal, for $R=1$ cm at a frequency of one kHz
    ($kR=0.2354$). The soft layers lay at $z_1=0.3 R$ and $z_2=0.9 R$, at the
    interfaces between slab 0, left, slab 1, middle, and slab 2, right. The
    dash-dot lines mark the discontinuities of $u(z)$. The dash-dash line
    represents the elastic behaviour with no soft layers.}
\end{figure}

\subsection{Soft layers}

To account for the effect of highly deformable dislocation structures, the
model is extended by introducing two soft layers at $z_1$ and $z_2$ parallel
to the rigid cell walls, as depicted in Fig.\,\ref{waveform}. The slabs of
solid helium that these soft layers delimit have the elastic properties of the
homogeneous crystal discussed in subsection \ref{HomogeneousCase}
above. 

Strain and stress are continuous functions at the interface between the slab
of dislocation-free crystal and the soft layer.  Denoting the shear modulus in
the soft layer $G_{\m s}$, shear plane waves propagate with wave vector $k_{\m
  s} = \sqrt{\rho/G_{\m s}}$. The propagation of the deformation-stress
vector, $[u(z),\sigma(z)]$, in the layer of thickness $d$ is described by the
transfer matrix
\begin{displaymath}
  \b M(d) = \left[
    \begin{array}{lr}
      \cos k_{\m s}d              &  (1/k_{\m s}G_{\m s}) \sin k_{\m s}d
    \\ -k_{\m s}G_{\m s} \sin k_{\m s}d  &      \cos k_{\m s}d
    \end{array}
    \right]  \; .
\end{displaymath} 
Although the problem of finding how plane waves propagate through the stack of
slabs depicted in Fig.\,\ref{waveform} is formally solved by multiplying
transfer matrices such as $\b M(d)$, it saves a number of algebraic steps to
let the thickness $d$ and the modulus $G_{\m s}$ go to zero in such a way that
$d/G_{\m s}$ remains constant and equal to $\alpha$. The effect of the soft
layer is then lumped into a discontinuous jump in the deformation proportional
to the local stress, described, for the layer at $z_1$, by
\begin{eqnarray}    \label{SoftLayerBoundaryConditions}
  u(z_1+d)      & = & u(z_1) + \alpha_1 \sigma(z_1) \; ,  \\
  \sigma(z_1+d) & = & -k^2_{\m s} d G_{\m s} u(z_1) + \sigma(z_1) 
       \nonumber  \; ,
\end{eqnarray} 
In the limit $d \rightarrow 0$, the last equality expresses the continuity of
stress across the infinitely thin layer while the displacement experiences a
discontinuity.  These boundary conditions, which could have been anticipated,
also apply to the soft layer at $z_2$ with slip parameter $\alpha_2$. In the
following, the soft layers will be described by their compliances
$\alpha_i/R$, which are such that the parameters $\kappa_i = \alpha_i G/R$ are
dimensionless quantities.

\section{Model analysis} \label{Results} 

\subsection{Wave propagation through the sample} \label{WaveTroughSample}

With the boundary conditions, Eqs.(\ref{SoftLayerBoundaryConditions}),
describing the soft layers, the propagation of the propagating and
counter-propagating waves through the three slabs of homogeneous crystal with
shear modulus $G$ and obeying no-slip boundary conditions at the walls can be
found by straightforward algebra.

Wave propagation in slab 0, between $z=0$ and $z_1$ as shown in
Fig.\,\ref{waveform}, is represented by Eq.(\ref{SimpleWaveEquationSolution}),
which involves two integration constants $\u0{}+$ and $\u0{}-$, the amplitudes
of the counter-propagating plane waves with pulsation $\omega$ and wavevectors
$\pm k$. Similar solutions obtain in slab 1 between $z_1$ and $z_2$, and in
slab 2 between $z_2$ and $z=R$, involving constants $u_{\ss 1+}$, $u_{\ss
  1-}$, and $u_{\ss 2+}$, $u_{\ss 2-}$ respectively.

The amplitudes of the propagating and counter-propagating waves in slab 2
are linearly related to those in slab 0:  
\begin{eqnarray}        \label{DefinitionDelta}
    u^{}_{\ss{2+}} & = & \delta^{}_{11} \u0{}+ +\delta^{}_{12} \u0{}-  \; , \\ 
    u^{}_{\ss{2-}} & = & \delta^{}_{21} \u0{}+ +\delta^{}_{22} \u0{}-  \; . \nonumber  
\end{eqnarray}
\begin{widetext}
The deformation discontinuity at the soft layer at $z_1$ yields the following
relations: 
\begin{eqnarray*}
    u_1(z_1) & = & u_{\ss{1+}} \m e^{\textstyle{-ikz_1}} +  
               u_{\ss{1-}} \m e^{\textstyle{ikz_1}}  
             = \u0{}+ (1-i \kappa_1 kR) \m e^{\textstyle{-ikz_1}} +
               \u0{}- (1+i \kappa_1 kR) \m e^{\textstyle{ikz_1}} \; , 
             \\
   \sigma(z_1) & = &  -ikG\left( \u0{}+  \m e^{\textstyle{-ikz_1}} -
               \u0{}- \m e^{\textstyle{ikz_1}} \right)
                = -ikG\left( u_{\ss{1+}} \m e^{\textstyle{-ikz_1}} -  
               u_{\ss{1-}} \m e^{\textstyle{ikz_1}} \right) \; .
\end{eqnarray*}
Similar relations hold between  $u_{\ss 1+}$, $u_{\ss 1-}$, and
$u_{\ss 2+}$, $u_{\ss 2-}$. Eliminating  $u_{\ss 1+}$ and $u_{\ss 1-}$ leads to
the following expressions for the coefficients $\delta_{ij}$ of the matrix
that describes wave propagation through the stack of slabs 0, 1, 2:
 \begin{subequations}       \label{CoefficientsDelta}
  \begin{eqnarray}  
    \delta_{11} & = & \left(1-i\frac{\kappa_1}{2} kR\right)
                     \left(1-i\frac{\kappa_2}{2} kR\right)
       +\frac{\kappa_1\kappa_2}{4}k^2R^2\; \m e^{\textstyle{2ik(z_2-z_1)}} 
       = \delta_{11}' + i\delta_{11}'' \; ,
    \label{delta11}\\
    \delta_{12} & = & i \frac{\kappa_1}{2} kR
         \left(1-i\frac{\kappa_2}{2} kR \right) \m e^{\textstyle{2ikz_1}}
                   + i \frac{\kappa_2}{2} kR
         \left(1+i\frac{\kappa_1}{2} kR \right) \m e^{\textstyle{2ikz_2}}
       = \delta_{12}' + i\delta_{12}''  \; ,
     \label{delta12}\\
    \delta_{21} & = & -i \frac{\kappa_1}{2} kR
         \left(1+i\frac{\kappa_2}{2} kR \right) \m e^{\textstyle{-2ikz_1}}
                     -i \frac{\kappa_2}{2} kR
         \left(1-i\frac{\kappa_1}{2} kR \right) \m e^{\textstyle{-2ikz_2}} 
       = \delta^\ast_{12} \; , 
     \label{delta21}\\
    \delta_{22} & = & \left(1+i\frac{\kappa_1}{2} kR\right)
                     \left(1+i\frac{\kappa_2}{2} kR\right)
       +\frac{\kappa_1\kappa_2}{4}k^2R^2\; \m e^{\textstyle{-2ik(z_2-z_1)}} 
       = \delta^\ast_{11} \; , 
    \label{delta22}
  \end{eqnarray}
 \end{subequations}
\end{widetext}
The matrix $\m \Delta = ||\delta_{ij}||$, Eq.(\ref{CoefficientsDelta}),
describing wave propagation in a conservative time-reversal invariant system,
is unitary and has determinant unity:
\begin{equation}        \label{UnitaryMatrix}
    \delta_{11} \delta_{22} -  \delta_{12} \delta_{21} = 1 \; .
\end{equation}

\subsection{Shear modulus} \label{SoftShearModulus}
 
For shear modulus measurements, the no-slip condition at the walls reads:
\begin{eqnarray*} 
  \u0{}+ + \u0{}- & = & \u0{}{} \; , \\
   u_{\ss 2+} \m e^{-ikR} + u_{\ss 2-} \m e^{ikR} & = & 0 \\
   =(\delta_{11}\m e^{-ikR} + \delta_{21} \m e^{ikR}) \u0{}+  & + &
          (\delta_{12}\m e^{-ikR} + \delta_{22} \m e^{ikR}) \u0{}- \; ,
\end{eqnarray*}
relations from which the integration constants  $\u0{}+$ and $\u0{}-$ can be derived:
\begin{eqnarray}            \label{ConstantsShear}     
   && \frac{\u0S+}{\delta_{12}\m  e^{-ikR} + \delta_{22}\m  e^{ikR}}
    =  \frac{-\u0S-}{\delta_{11}\m  e^{-ikR} + \delta_{21}\m  e^{ikR}} 
   \nonumber \\
   && = \frac{\u0{ }{ }}{(\delta_{12}-\delta_{11}) \m e^{-ikR} 
      - (\delta_{21}-\delta_{22}) \m e^{ikR}} =\widetilde{\u0{}{}}
  \; .
 \end{eqnarray}
The quantities $\u0S+$ and $\u0S-$ expressed by Eqs.(\ref{ConstantsShear})
now include the effect of the soft layers and should not be confused with
those given by Eq.(\ref{SimpleConstantsShear}), which do not. The shear
stress at the receiver $\sigma(R)$ is given by
\begin{displaymath}
    \sigma(R) = G \frac{\m d u}{\m d z}\Big|_\m R 
  = -ikG \big ( u_{\ss 2+} \m e^{-ikR} - u_{\ss 2-} \m e^{ikR} \big) \; .
\end{displaymath}
Expressing $ u_{\ss 2+}$ and $u_{\ss 2-}$ in terms of $\u0S+$ and $\u0S-$
using Eqs.(\ref{CoefficientsDelta}) and (\ref{UnitaryMatrix}),  $\sigma(R)$
takes the following simple form:
\begin{equation}            \label{StressShear}  
  \sigma(R) =- 2 ikG \widetilde{\u0{}{}}
   = -kG\u0{}{}/\cal{D} \; ,
\end{equation}
\begin{eqnarray*}   
  \text{with }  {\cal D}&=& (1/2i)\left[(\delta_{12}-\delta_{11}) \m e^{-ikR}
       -(\delta_{21}-\delta_{22}) \m e^{ikR}\right] \\
  &=& \sin kR + \kappa_1 kR \cos k(R-z_1) \cos kz_1 \\
  &+&\kappa_2 kR\cos k(R-z_2) \cos kz_2 \\
  &-&\kappa_1\kappa_2 k^2R^2 \sin k(z_2-z_1)\cos k(R-z_2) \cos kz_1 
   \nonumber \; . 
\end{eqnarray*}
The effective shear modulus $G_{\m{eff}} = -\sigma(R)\,R/\u0{}{}$ follows
readily from Eq.(\ref{StressShear}): 
\begin{subequations}
\begin{eqnarray}        \label{EffectiveShearModulus}
  \frac{G}{G_{\m{eff}}} &=& \frac{\cal D}{kR} \\
  &\simeq&1 +\kappa_1+\kappa_2
  -\frac{k^2R^2}{6}-\frac{\kappa_1+\kappa_2}{2}k^2R^2 \nonumber \\
  &+&\kappa_1 k^2z_1(R-z_1)+ \kappa_2 k^2z_2(R-z_2)\nonumber \\
  &-&\kappa_1\kappa_2 k^2R\,(z_2-z_1)+\cdots\; .
  \label{ExpansionShearModulus} 
\end{eqnarray}
\end{subequations}
For typical numerical values, such as those used for the graphs in
Fig.\,\ref{waveform}, the first two terms of the expansion of $1/G_{\m{eff}}$
with respect to $kR$, Eq.(\ref{ExpansionShearModulus}), fall within 1\% of the
exact value given by Eq.(\ref{EffectiveShearModulus}). The zeroth order
term could have been written from scratch. The correction to
the inertial mass turns out to be less transparent and is considered in the
next subsection.
\begin{figure}[t]
  \includegraphics[width=80mm]{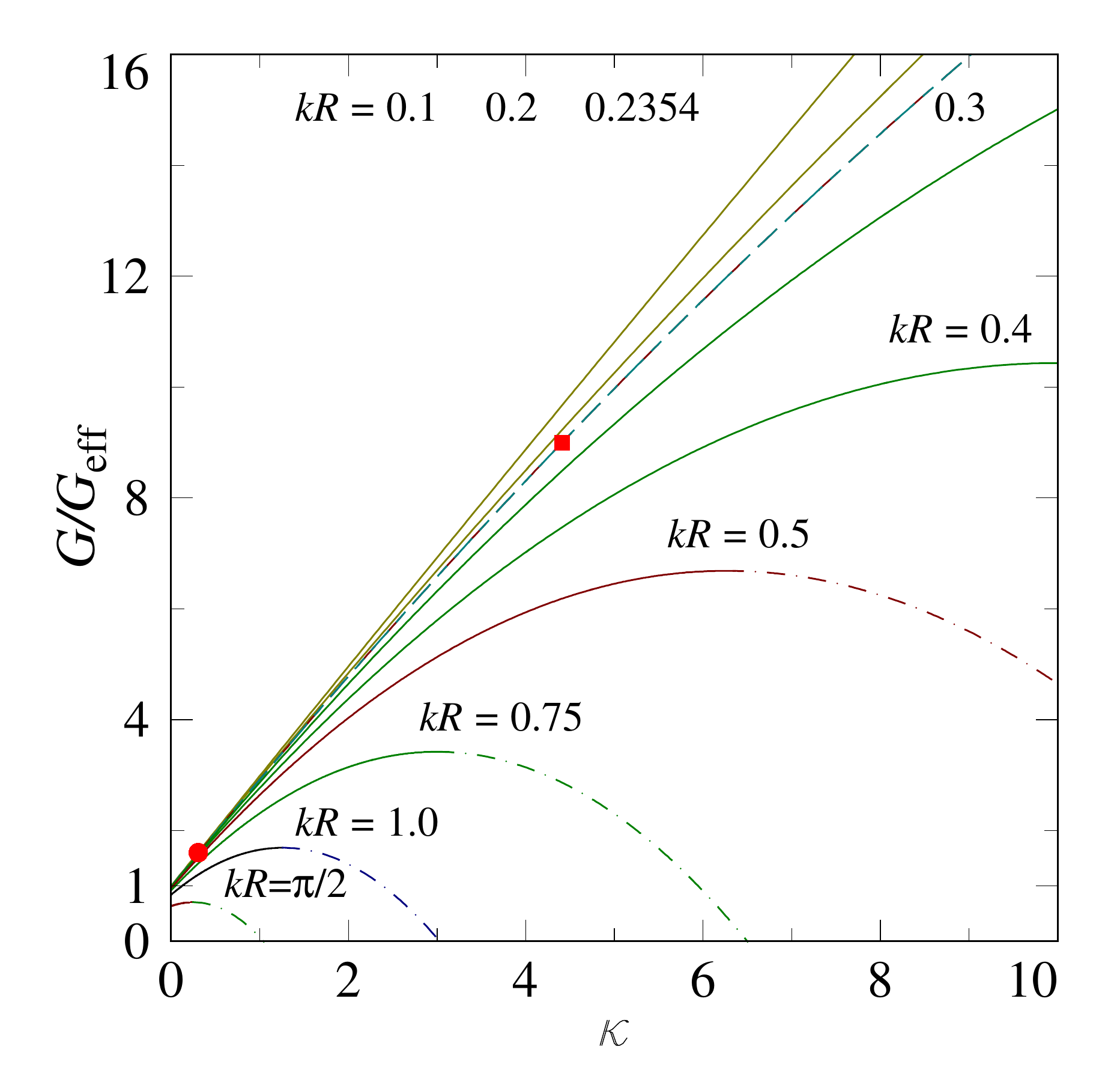}
  \caption{\label{Shear-vs-kappa}Inverse effective shear modulus
    $G_{\m{eff}}$, normalised to the shear modulus with no soft layers $G$ in
    terms of the dimensionless compliance $\kappa$ of the soft layers, for the
    same parameter values as in Fig.\,\ref{waveform} for various values of
    $kR$. The dash-dash curve represents the case with $R=1$ cm at a frequency
    of one kHz, $kR=0.2354$. The symbols mark the cases discussed in the text,
    ($\scriptscriptstyle\blacksquare$) for $G/G_{\m{eff}}= 9\;, \kappa=4.411$,
    ($\bullet$) for $G/G_{\m{eff}}= 1.6\;, \kappa=0.3075$. The dash-dot-dash
    lines correspond to a reentrant branch that, for a given value of $G/G_{\m{eff}}$,
    cannot be reached by adiabatic turn-on of the $\kappa_i$'s.  }
\end{figure}

\subsection{Effective mass}  \label{EffectiveMass}

The derivation of the apparent inertia of the
sample follows that given in Sec.\,\ref{HomogeneousCase} for the homogeneous
sample, starting from the same boundary conditions,
Eqs.(\ref{InertiaBoundaryConditions}). Equation (\ref{SimpleConstantsInertia})
for the integration constants is  modified as:
\begin{eqnarray}        \label{ConstantsSoftInertia}
   && \frac{\u0M+}{\delta_{12}\m  e^{-ikR} + \delta_{22}\m  e^{ikR}-1} =
   \nonumber \\
   &&   \frac{-\u0M-}{\delta_{11}\m  e^{-ikR} + \delta_{21}\m  e^{ikR}-1} 
   =\widetilde{\u0{}{}}
  \; ,
\end{eqnarray}
the quantity  $\widetilde{\u0{}{}}$ being the same as in
Eq.(\ref{ConstantsShear}) for the shear modulus case. 

Using these integration constants,  the stress at each wall takes the
following form:
\begin{eqnarray}        \label{SoftStress}
   \sigma(0) &=& G \frac{\m d u}{\m d z}\Big |_{z=0} 
      = - ikG(\u0M+ - \u0M-)   \\
   &=& ikG\widetilde{\u0{}{}}\,\Big[2-(\delta_{12}+\delta_{11})
      \m e^{-ikR} - (\delta_{22}+\delta_{21}) \m e^{ikR} \Big]
    \; , \nonumber \\
   \sigma(R)&=& G \frac{\m d u}{\m d z}\Big |_{z=R} 
      = - ikG\Big(u_{\ss 2+} \m  e^{-ikR} -u_{\ss 2-} \m  e^{ikR}\Big) \nonumber \\ 
   &=& -ikG\widetilde{\u0{}{}}\,
      \Big[2(\delta_{11}\delta_{22}-\delta_{12}\delta_{21})  \nonumber \\
   && \mbox{\hskip 4ex} -(\delta_{11}-\delta_{12}) \m e^{-ikR} 
     - (\delta_{22}-\delta_{21}) \m e^{ikR} \Big] \; . 
\end{eqnarray}
The total force per unit area exerted by the helium sample on both walls is now
given, instead of Eq.(\ref{SimpleInertia}) for the homogeneous case, by
\begin{displaymath}       
  F_{\m X}+F_{\m R}=\sigma(0)-\sigma(R) = 2ikG\widetilde{\u0{}{}}\,
      \Big[2 - \delta_{11} \m e^{-ikR} - \delta_{22} \m e^{ikR}\Big] \; ,
\end{displaymath}  
using again the property that $||\delta_{ij}||$ has determinant
unity. Expliciting the quantities within square brackets making use of
Eqs.(\ref{CoefficientsDelta}) and Eq.(\ref{EffectiveShearModulus}), the total
force on the walls takes the final form
\begin{equation}        \label{SoftInertia}
  F_{\m T} = F_{\m X}+F_{\m R}= kGu_{\ss 0}\frac{\cal N}{\cal D} 
          =\frac{u_{\ss 0}}{R} G_{\m{eff}} {\cal N} \; ,
\end{equation}
\begin{eqnarray*}       
  \text{with } {\cal N} &=& 2 - \delta_{11} \m e^{-ikR} - \delta_{22} \m e^{ikR} 
    = 2(1-\cos kR) \\
    &+& (\kappa_1+\kappa_2)kR \sin kR 
     - \kappa_1\kappa_2 k^2R^2\sin k(z_2-z_1) \\
    &\times& \big\{\sin k(z_2-z_1)+\sin k(R-z_2+z_1) \big\} \; . 
\end{eqnarray*}

\mbox{} 
\subsection{Stationary waveforms}  \label{waveforms}

The displacement $u(z)$ in the sample can easily be evaluated using, {\it
  e.g.} in the inertia measurement case, the solution to the wave equation
expressed by Eqs.(\ref{ConstantsSoftInertia}) for $\u0M+$ and $\u0M+$, with
the following result:

\begin{itemize}
  \item{for slab 0:
  $ u(z)^{(0)} = u_0[\cos{kz} +(B/A) \sin{kz})] \; ; $
  } 
  \item{for slab 1:
  $ u(z)^{(1)} =u_0[\cos{kz} - \kappa_1kR/2\{\sin kz+\sin k(2z_1-z)\}
    +(B/A)[\sin kz+ \kappa_1kR/2\{\cos kz + \cos k(2z_1-z)\}]]\; ; $
  }
  \item{for slab 2:
  $ u(z)^{(2)} =(u_0/A)[\sin k(R-z)+\sin kR+ \kappa_1kR/2 \cos k(z-z_1) 
    \cos kz_1 + \kappa_2kR/2 \cos k(z-z_2) \cos kz_2 -
    \kappa_1\kappa_2 K^2G^2 \sin k(z_2-z_1) \cos k(z-z_2) \cos kz_1] \; .
  $
  }
\end{itemize}  
In these expressions, 
\begin{eqnarray*}       
  A &=& ( \delta_{11}'- \delta_{12}')\sin kR +  
        ( \delta_{12}''-\delta_{11}'')\cos kR = \cal D \;,\\
  B &=& 1-(\delta_{11}'+ \delta_{12}')\cos kR 
         -(\delta_{12}''+\delta_{11}'')\sin kR \;  .
\end{eqnarray*} 
These waveforms can readily be evaluated numerically. As an example, the
relative displacement $u(z)/u_0$ for three values of the dimensionless
compliance $\kappa$ of the soft layers, taken to be equal, is shown in
Fig.\,\ref{waveform} for $kR=0.2354$, $z_1=0.3 R$ and $z_2=0.9 R$. The
discontinuities at $z_1$ and $z_2$ caused by these soft layers increase in
size with the compliance, up to the point where $\cal D$ becomes zero and the
deformation diverges.

The next step, carried out in the following Section, consists in extracting
the parameters $\kappa_i$ of the soft layers from the measured value of
$G_{\m{eff}}$ and in evaluating the corresponding apparent change in inertia.

\begin{figure}[t]
  \includegraphics[width=80mm]{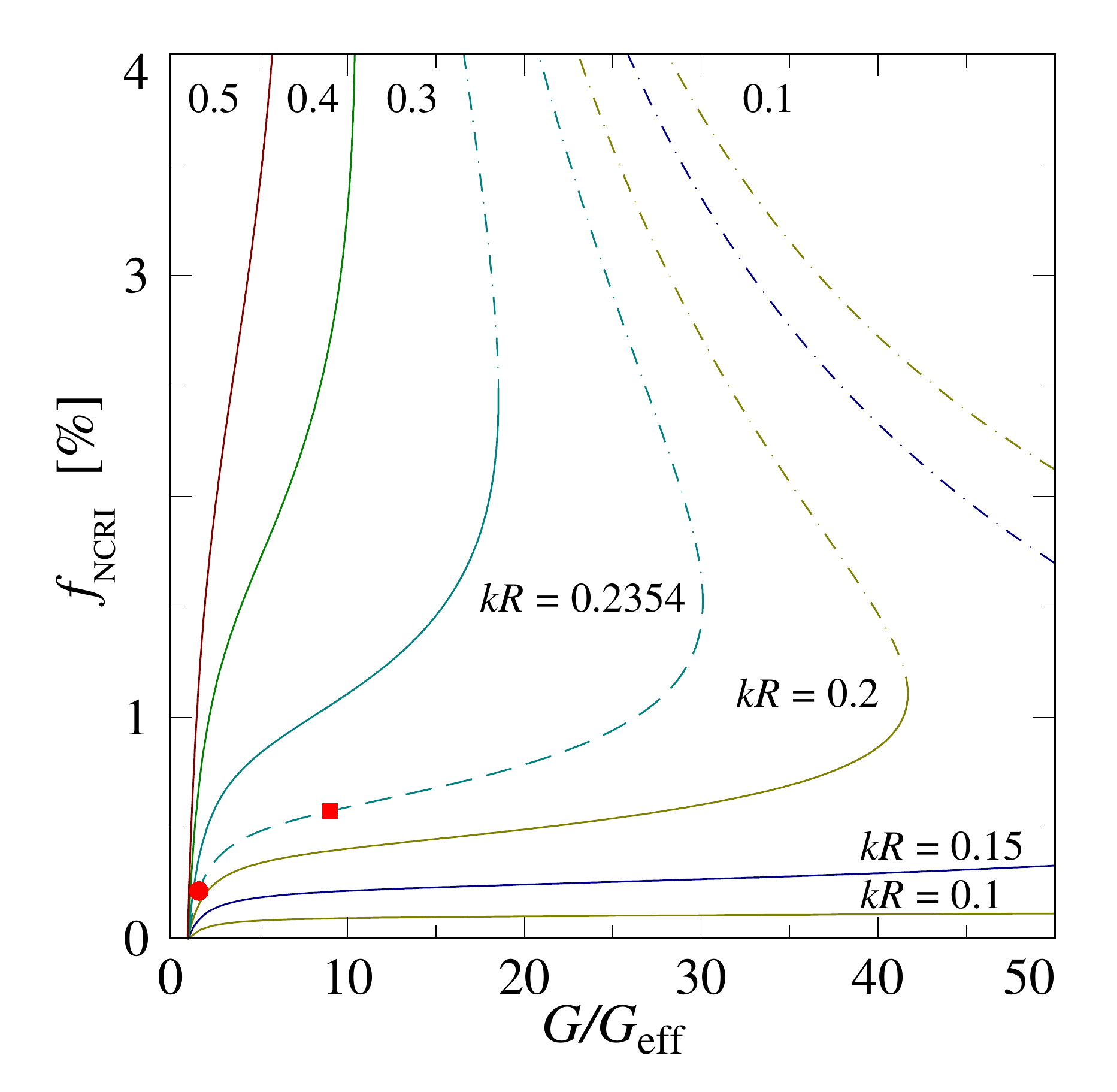}
  \caption{\label{Inertia-vs-Shear}Relative change in the apparent inertia,
    $f_{\m{NCRI}}$ vs the inverse dimensionless shear modulus $G/G_{\m{eff}}$,
    for the situation of Fig.\,(\ref{waveform}), for various values of $kR$ as
    labelled in the figure. The symbols ($\bullet$) and
    ($\scriptscriptstyle\blacksquare$) on the dash-dash curve for $kR=0.2354$
    mark the same cases as in Fig.\,\ref{Shear-vs-kappa}. The dash-dot-dash
    portions of the various curves correspond to reentrant regions that are
    not be reached by adiabatic turn-on of the compliance of the soft layers.}
\end{figure}

\section{Numerical results} \label{Discussion} 

As the temperature is raised from absolute zero, $\kappa$ varies from its low
temperature value, assumed to be negligibly small because the dislocations are
immobilised by the isotopic impurities, to its high $T$ value. The
corresponding change of $G/G_{\m{eff}}$ to lowest order in the small parameter
$kR$ in Eq.(\ref{ExpansionShearModulus}) reduces to a static correction to the
elastic constant. The lowest order correction to the
effective mass results from a dynamical effect of a magnitude
comparable to that of the plain elastic response, which should be subtracted
out. This difference follows from Eqs.(\ref{SimpleInertia}) and
(\ref{SoftInertia}):

\begin{subequations}    
\begin{eqnarray}   \label{MassCorrection}      
  \hskip -0.7cm 
    F_{\m T}-F_{\m T}\big |_{\kappa_1,\,\kappa_2=0} &=                       
    \frac{\rho R\omega^2 u_{\ss 0}}{kR} \Big \{
    \frac{\cal N}{\cal D} -2\frac{1-\cos kR}{\sin kR}\Big \}\;\; 
     \label{FullMassCorrection} \\  
   &\simeq \rho R\omega^2  u_{\ss 0} \,
   \frac{k^2R^2}{1+\kappa_1+\kappa_2}\, \times 
   \bigg[\frac{\kappa_1+\kappa_2}{4}    \nonumber \\
   &- \kappa_1\frac{z_1(R-z_1)}{R^2} - \kappa_2\frac{z_2(R-z_2)}{R^2} 
     \bigg] \; .  \label{MassCorrectionExpansion}
\end{eqnarray} 
\end{subequations}   
Equations (\ref{MassCorrection}) and (\ref{MassCorrectionExpansion}) show how
what could be called the ``superfluid fraction'', $f_{\m{NCRI}} = (F_{\m T}-F_{\m T}
|_{\kappa_1,\,\kappa_2=0})/\rho R \omega^2 u_{\ss 0}$ depends
on the compliances $\kappa_i$, which in turn are related to the effective
shear modulus.  The quantity $\rho R\omega^2 u_{\ss 0}$ has already appeared
in Eq.(\ref{SimpleInertia}) and stands for the force due to the acceleration
of the inertial mass $M_{\m I}=\rho R$. These quantities hold per unit area.

The full expression of the exact result, Eq.(\ref{FullMassCorrection}), is
fairly lengthy and not particularly transparent but evaluates numerically
quite readily. The outcome is discussed below.  The lowest order correction to
this effective mass, Eq.(\ref{MassCorrectionExpansion}), is second order in
$kR$ and linear in the $\kappa_i$'s, the term in $\kappa_1\kappa_2$, of order
$(kR)^3$, being discarded. This correction is either positive or equal to zero
for the special case $z_1=z_2=R/2$, that is, for a vanishing
dangling mass, and, by symmetry, vanishing local stress in slab 2.

\medskip Equations (\ref{MassCorrection}) and (\ref{MassCorrectionExpansion}),
together with (\ref{EffectiveShearModulus}) and (\ref{ExpansionShearModulus}),
which express the NCRI fraction and the effective shear modulus in the
presence of soft dislocation arrays, constitute the main result of this
work.\cite{PostonCondMat,however}

\medskip
The variation of the shear modulus in terms of the soft layer compliances,
taken for simplicity to be equal to a common value $\kappa$, is shown in
Fig.\,\ref{Shear-vs-kappa} for various values of $kR$ for the same sample
geometry and parameter values as in Fig.\,\ref{waveform}.  As the compliance
$\kappa$ increases from zero, assumed to be its $T=0$ value, the effective
shear modulus $G_{\m{eff}}$ decreases; the solid becomes softer, up to a point
where $G/G_{\m{eff}}$ reaches a maximum: the interfaces between slab 1 and its
neighbours becomes so soft that, although the dangling slab swings with
increasing amplitude, the stress due to its motion ceases to increase. Beyond
this point, a further increase in $\kappa$ would lower $G/G_{\m{eff}}$ because
the stress reflected back onto the external boundaries effectively decreases
while the displacement of slab 1 goes on increasing.

It has been assumed above that the steady-state regime is reached
adiabatically, which implies: 1) that only the ascending branch of
$G/G_{\m{eff}}$ in Fig.\,\ref{Shear-vs-kappa} can be reached by adiabatic
turn-on of the $\kappa_i$'s from zero; 2) that damping does not vanish
entirely. If damping is introduced in the wave equation,
Eq.(\ref{SimpleWaveEquation}), slab 1 would be coupled to its neighbours by
friction in addition to shear elasticity and the results obtained above would
be quantitatively different from those in Fig.\,\ref{Shear-vs-kappa} for very
small values of $G_{\m{eff}}$. In particular, the descending branch of
$G/G_{\m{eff}}(\kappa)$ could not actually cross the $x$-axis.

The NCRI fraction is plotted directly in terms of the shear modulus in
Fig.\,\ref{Inertia-vs-Shear}.  From the measured overall change in
$G/G_{\m{eff}}$, which reaches values of
1.6~\cite{Tsymbalenko:79,Mukharsky:09} up to 9 or more,\cite{Rojas:10} the
corresponding values of the compliance of the soft layers can be found from
Eq.(\ref{EffectiveShearModulus}). From these values, $\kappa=0.31$ for
$G/G_{\m{eff}}=1.6$, $\kappa=4.4$ for $G/G_{\m{eff}}=9$, the NCRI fractions
given by Eq.(\ref{MassCorrection}) are 0.22 \% and 0.58 \% respectively. These
values depend on $z_1$ and $z_2$: the largest $f_{\m NCRI}$ are achieved for
$z_1\sim 0$, $z_2\sim R$. The reentrant branches of the graphs in
Fig.\,\ref{Inertia-vs-Shear} correspond to the descending branches for large
$\kappa$ in Fig.\,\ref{Shear-vs-kappa}. An accurate description of these regimes
where displacements become very large should, as already mentioned, include
damping. They are irrelevant to the present discussion.

\begin{figure}[t]
  \includegraphics[width=80mm]{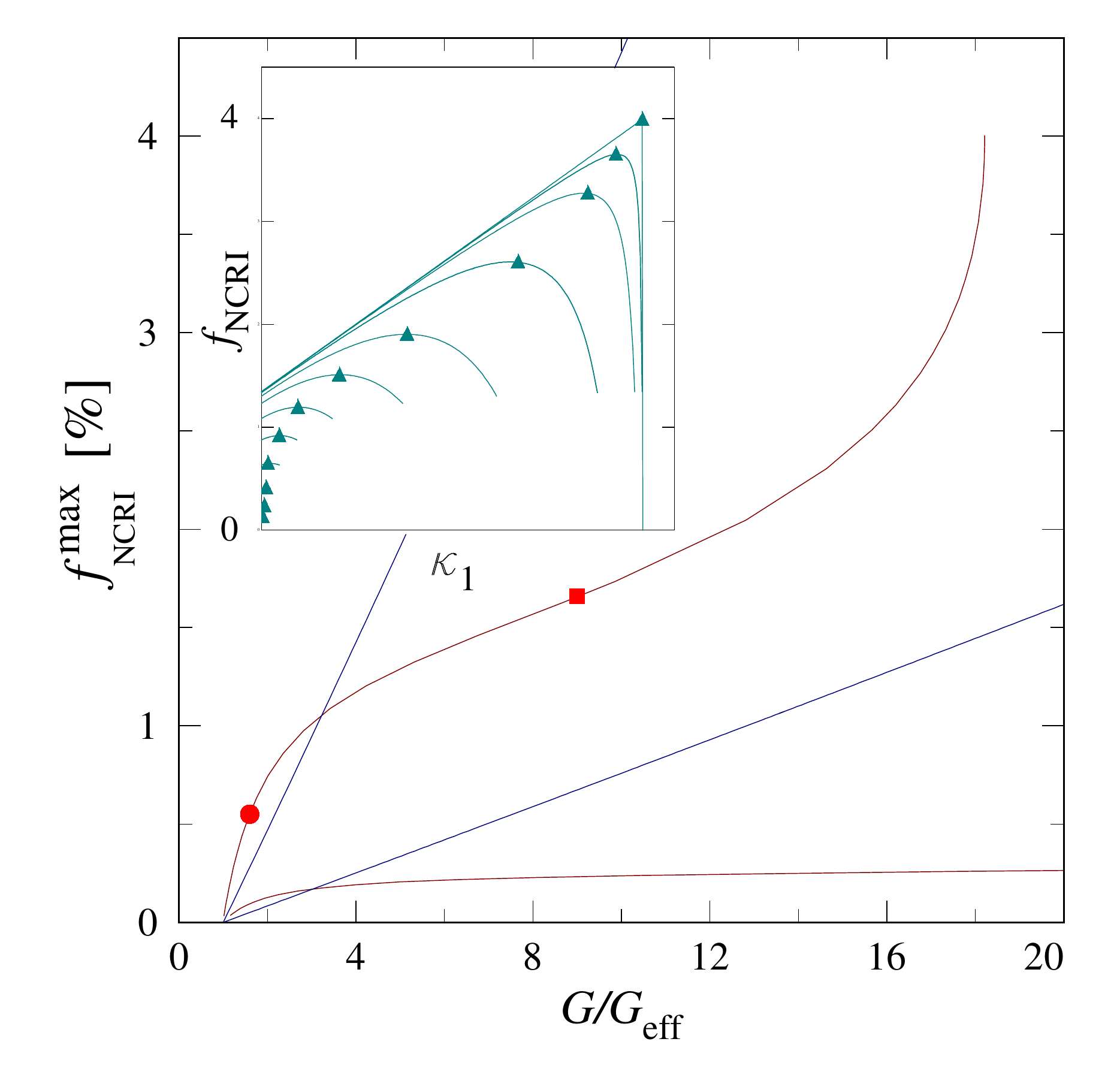}
  \caption{\label{InertiaExtremum} Maximum possible value of $f_{\m{NCRI}}$ in
    terms of $G/G_{\m{eff}}$ for $kR$=0.2354 (upper curves) and 0.1 (lower
    curves). The nearly straight curves stem from the homogeneous elastic
    response, the bending curves are the pure soft layer contributions. The
    values for $f_{\m{NCRI}}^{\m{max}}$ are extracted from graphs for
    $f_{\m{NCRI}}$ in terms of $\kappa_1$ as shown in the insert where they
    are marked by ($\blacktriangle$). The symbols ($\bullet$) and
    ($\scriptscriptstyle\blacksquare$) on the curve for $kR=0.2354$
    mark the same cases as in Fig.\,\ref{Shear-vs-kappa}. }
\end{figure}

Dislocations are also found in the homogeneous slabs. They may also induce
a variation of $k=\sqrt{\rho/G}\,\omega$ as $G$ may also vary with temperature. The
contributions to $f_{\m{NCRI}}$ of the soft layers and of the dislocation
network in the homogeneous slabs are seen in Eq.(\ref{MassCorrection}) to be
additive and their respective weights depend on how each contributes to
$G/G_{\m{eff}}$. However, the contribution of the network should be no more
than 20\%, the geometrical limit for hcp structures, so that its effect on
$f_{\m{NCRI}}$ is less significant.

The highest values of $f_{\m{NCRI}}$ for given $G/G_{\m{eff}}$ and $kR$ are
reached for $z_1=0$, $z_2=R$ and varying $\kappa_1$ while adjusting $\kappa_2$
to keep $G/G_{\m{eff}}$ constant. These {\it maximum maximorum} values are
plotted in Fig.\,\ref{InertiaExtremum} for $kR=0.2354$ and 0.1, and for
various values of the effective shear modulus. These quantities overtake those
for homogeneous systems, given by Eqs.(\ref{SimpleEffectiveModulus}) and
(\ref{SimpleInertia}), up to $G/G_{\m{eff}}$ values that are much larger than
those for homogeneous systems $\Lambda L_{\m{N}}^2\leq 2^{-1/2}$, as seen in
Fig.\,\ref{InertiaExtremum}. Values of $f_{\m{NCRI}}$ ranging from near zero
to above one per cent can be reached for $kR=0.2354$, {\it i.e.} in a one cm
size cell at a frequency of 1 kHZ for the experimentally observed values of
$G/G_{\m{eff}}$.  For a cylindrical geometry, these values are approximately
halved.\cite{Clark:08}

\section{Discussion} \label{Conclusion} 
Because of the very large drop of the shear modulus observed at low
frequencies in most samples of solid $^4$He, it is surmised in this work that,
instead of forming homogeneous random networks, dislocations crop into
organised slip bands or quasi-planar arrays of sizable thickness.  When the
$^3$He impurities evaporate from the cores of the dislocation lines, the
latter become very mobile and the planar arrays very soft.\cite{Amodeo:90} It
is argued that a large degree of polycrystallinity does not suffice to obtain
large $G/G_{\m{eff}}$ and large $f_{\m{NCRI}}$, as shown experimentally in
aerogel by \citet{Mulders:08}. Dislocations have to organise over large
distances in such a way that parts of the sample become uncoupled and acquire
additional kinetic energy, thereby increasing the apparent inertia.

In a hcp structure, edge dislocations glide easily in the basal plane along
three preferred crystallographic directions at 120 degrees of one another. To
climb away from these directions, they have to change into screw or mixed
dislocations. As shown by \citet{Suzuki:85} this process is thermally assisted
above $\sim$0.8~K (at $\rho =0.192$~g/cm$^3$) and proceeds by
quantum-tunnelling below. More recent and detailed theoretical considerations
of the climb process in the quantum regime\cite{Shevchenko:87,Boninsegni:07}
have led to the realisation that the cores of screw and edge dislocations
could become superfluid. Climb processes would then become greatly enhanced,
hence the concept of ``superclimb'' introduced by Kuklov and
coworkers.\cite{Aleinikava:08,Soyler:09,Aleinikava:10,Aleinikava:11} This
quantum-assisted climb process provides a mechanism in solid $^4$He for edge
dislocations to easily move off the basal plane. This process lifts a
constraint on dislocation motion. The propagation of dislocation pile-ups in
the course of plastic deformation becomes greatly facilitated, as well as the
formation of percolating planar defects.

On heuristic grounds, propagation of dislocation pile-ups in brittle
materials, such as hcp helium, causes cracks to form and results eventually in
mechanical failure. In solid helium, which is under positive hydrostatic
pressure, cracks with voids cannot form but corresponding macroscopic defects
with little or no crystalline order must appear.\cite{Takeuchi:76,Amodeo:90}
Hence the plausible appearance of connected veins imprinted by plastic
flow. These regions of the sample show strong spatial disorder and can
possibly sustain off-diagonal long range order
instead.\cite{Penrose:56,Boninsegni:06} That they are found anomalously soft
in a number of experiments lends credence to this possibility.

The model based on these soft layers is easily tractable analytically. The
calculated values for the shear modulus and the NCRIf fall within the range of
observed values, barring the highest ones.\cite{Rittner:07} This model
explains readily why the NCRI and stress-strain measurements depend so
strongly on the sample geometry\cite{Rittner:07} and thermal
history:\cite{Mukharsky:07,Clark:08} even small changes in the soft layer
properties and the interconnection of the channels that they delimit can
greatly influence the motion of the dangling masses. Homogeneous dislocation
networks, besides having a limited effect of the shear modulus, can hardly
exhibit such variability.

Actual samples studied in the laboratory are likely to be more convoluted than
sketched in Fig.\,\ref{waveform}. The veins have tortuous paths and coarse
sheaths, which might appear to hinder motion. However, applied strains are
small, of the order of $10^{-6}$ or
less,\cite{Syshchenko:09,Mukharsky:09,Day:10} and displacements are
correspondingly small.  The soft layers considered here are thicker than
low-tilt angle subgrain boundaries, possibly in the 10 to 100 nm
range.\cite{Amodeo:90} The crystal lattice is heavily distorted over such a
thickness.  The soft layers can be expected to be quite malleable on such a
scale and yield easily under local stress. Being extremely compliant, they
support plastic flow within themselves and accommodate departures from the
ideal planar geometry depicted in Fig.\,\ref{waveform}.  The soft layers can
conceivably also become fully fluid,\cite{Suzuki:73} or even genuinely
superfluid as already mentioned above.\cite{Boninsegni:06} Crystal subgrain
motions on a sub-millimetric scale have actually been reported by \citet{Burns:08}
in X-ray experiments using very fine collimated beams. Similarly, the mobile
features observed in solid $^4$He at higher temperatures
\cite{Eyal:10,Eyal:11,Ray:09Comment} can be re-examined in the present
framework; these experiments also provide possible clues for the existence of
veins of easy deformation.

Maris and Balibar \cite{Maris:10} take a quite different point of view to
account for the observed relationship between $G_{\m{eff}}$ and
$f_{\m{NCRI}}$. They point out that experimental TO's may lack sufficient
structural rigidity. If the TO body deforms in such a way as to induce
additional strain on the helium sample, the NCRIf may appear larger than the
intrinsic value. As discussed in Ref.[\onlinecite{Maris:10}], the effect can
be quite large. This helps in particular to understand some very large NCRIf
values reported in the literature\cite{Rittner:08,Mi:11} that would not be
readily explainable with the existence of soft layers as assumed here. From
the results in Figs.\ref{Inertia-vs-Shear} and \ref{InertiaExtremum},
$f_{\m{NCRI}}$ remains below a few percent at the most for centimetre size
cells, less for smaller toroidal annuli. But conversely, these results do not
imply that, whenever the stiffness of the helium sample changes, an apparent
NCRI is bound to occur; this last feature depends on the geometry of the soft
layers and may be altogether absent.\cite{Fefferman:11}

A number of experiments might seem to invalidate the present approach. The TO
experiments with a blocked channel show a much reduced NCRIf. This is
interpreted as the manifestation that some sort of superflow is taking place
when flow paths are connected in a loop and not when the loop is
broken. However, the same considerations apply to the plastic flow in
connected veins, which also can form, or not, channels through which dangling
masses can jiggle.

Torsional oscillator measurements in confined geometries, Vycor, porous gold,
aerogel,\ldots, do show a sizable NCRIf and would also appear to completely
invalidate the present approach. If the model is applied to a single pore, for
which $kR$ is very small, then, indeed, the resulting effect that decreases as
$(kR)^2$ will be extremely small.  For sizes comparable to that of the soft
layers, the soft layer model is not expected to apply, neither for shear nor
for inertia. Pores do not appear to be filled with homogeneous hcp solid but
with either a combination of layers of liquid and of bcc
solid\cite{Wallacher:05} on top of 1$\sim$2 layers of amorphous solid, or, for
finer pores (47 \AA\ in MCM-41 and 34 \AA\ in gelsil),\cite{Bossy:10} with
amorphous solid only and inclusions of bcc-like nodules.  What was assumed for
softer layers carries over to the fine pores, which present a multi-connected
geometry with complex plastic flow patterns. The conditions of existence of
connected veins assumed at a macroscopic level are clearly fulfilled at the
mesoscopic level of the pores so that helium, either liquid or amorphous,
would contribute to NCRI. These questions deserve further
consideration.\cite{Maris:10,Reppy-private-communication}

Hexagonal solid $^3$He is also soft but appears not to show NCRI: the two isotopes
apparently possess similar elastic properties but different inertial
properties.  This isotopic dependence is well documented, in particular by the
work of \citet{West:09}. This observation would seem to also invalidate the
present approach. However, the tunnelling motion of dislocations is unlikely
to proceed in a similar manner in the bosonic and fermionic solids. In
particular, the process of superclimb,
\cite{Aleinikava:08,Soyler:09,Aleinikava:10,Aleinikava:11} which may assist
the formation of connected plastic flow veins, relies on the existence of
superfluidity in dislocation cores.\cite{Shevchenko:87,Boninsegni:07} This
mechanism does not operate in solid $^3$He.

\citet{Kim:11} directly addressed the connexion between shear and
NCRI in an ingenious experimental arrangement allowing simultaneous
measurements of both quantities. They observe, in particular, that the
response to an increase in drive amplitude differs very significantly between
both properties. However, the drive is not applied in an identical manner for both
measurements because of details of the cell geometry. Soft layers can be located
at different places and have different conformations: they are bound to 
respond differently.

Specific experiments can be performed to probe the present model. Shear modulus
measurements have not been performed in a cell geometry for which the plastic
flow lines would close on themselves in the way they do in torsional
oscillators. These measurements should reveal the existence of supersoft
elastic moduli.\cite{Ray:09Comment} Equipping a torsional oscillator with a
floppy membrane as {\it septum} to interrupt a quantum-coherent flow but not
the continuity of stresses and strains offers another venue.\cite{Mukharsky1}
The study of higher resonance modes in multiply-connected acoustic cavities
can also provide a way to probe the internal response of an inhomogeneous
sample.\cite{Mukharsky:12}

Multiple-mode TO resonators\cite{Aoki:07,Aoki:08,Mi:11} appear to give
somewhat indecisive answers but still show the expected trend of enhanced
NCRIf at higher frequency.\cite{Reppy-private-communication} A two-mode TO
with the dummy massive bob {\it inside} the resonator chamber, in contact with
the solid helium but connected loosely to the main body by an additional
torsion rod provides a mean of coupling shear to the sample in a Couette-type
experiment. If the inner bob angular position could be tracked by some optical
or electrodynamical means, the coupled system response could be analysed in
detail. A strong enhancement of $G/G_{\m{eff}}$ is expected, which would be
directly related to the NCRIf. A control experiment with bcc $^3$He, which
shows no shear modulus anomaly and no NCRI,\cite{West:09} can be carried out
at appropriate density and shear modulus values to distinguish between cell
and sample contributions to the apparent NCRI.

To conclude, the soft layer model presented here takes into account known
heterogeneities in dislocation patterns revealed in particular by the
anomalous softening of most samples of hcp $^4$He. It is argued that the
actual softness can be even more extreme than observed, being hampered by the
tortuous arrangement of the dislocation structures and of the crystalline
regions that they delimit. The corresponding values of the NCRIf are shown to
lie within the range of observations, barring the highest ones. The model
conflicts in no irredeemable way with the available assortment of experimental
observations. Conversely, it can be stated that most existing experiments to
date support the assumption of the existence of very mobile macroscopic veins
arranged along connected paths in hcp crystals of helium 4 and formed in a
process that depends on quantum statistics, like superclimb. Matter in the
veins themselves undergoes displacements governed by classical mechanics and
subject to dissipative mechanisms. These various assertions are amenable to
experimental verification.

\begin{acknowledgments}
  The author acknowledges useful discussions with Izumi Iwasa, S{\'e}bastien
  Balibar and Yuri Mukharsky and correspondence with John Reppy. He thanks
  Alan Braslau for his numerous suggestions on the manuscript. This work has
  been supported by ANR grant ``Superdur''.
\end{acknowledgments}

\bibliography{PRB11} 

\begin{thebibliography}{76}
\bibitem [{how()}]{however}%
  \BibitemOpen
  \href@noop {} {}\bibinfo {note} {The relation between $f_{\m{NCRI}}$ and
  $G_{\m{eff}}$ reported in Ref.[\onlinecite{Iwasa:10}] should stem from the
  present Eqs.(\ref{SimpleEffectiveModulus}) and (\ref{SimpleInertia}) for the
  homogeneous sample with $G$ replaced by $G_{\m{eff}}$. This corrected result
  is obtained directly by applying the boundary condition at the wall of the
  torsional oscillator to the full displacement $u(r,t)$ instead of just to the
  elastic part $u_{el}(r,t)$ in Eq.(22) of Ref.[\onlinecite{Iwasa:10}]. The
  same modification holds in Eq.(23) and leads to the cancellation of first
  order correction in $kr$ to the TO period. The second order term is that
  obtained by \citet{Clark:08} for a cylindrical geometry.}\BibitemShut {Stop}%
\end{thebibliography}
\appendix*
\section{ }
In a crystal lattice with hexagonal symmetry, there exists three glide
directions for edge dislocations in the basal plane, perpendicular to the
$\hat{\bf c}$-axis, at $2\pi/3$ from one another. These dislocations arrange
themselves on a hexagonal network in the basal plane with side length $a$,
possibly connected to adjacent basal planes at a distance $c$ along the
$\hat{\bf c}$-axis by pillars of screw or mixed dislocations.

The dislocation network that entirely fills a given basal plane of a sample
taken as a square of side $S$ for simplicity, can be constructed as shown in
Fig.\,\ref{hcp-dislocation}. The building block in thicker line is duplicated
and translated by $\vec A$ along one side and $\vec b$ along the other. There
are a total of $S^2/bA$ such translations, each involving a dislocation line
length equal to $6a$. The volume spanned in the process is $S^2c$ so that the
density of edge dislocation amounts to $\Lambda=6a/bAc$. As $b=3^{1/2}a$ and
$A=3a$, there comes that $\Lambda a c = 2/3^{1/2}$. This result has already
been quoted by \citet{Iwasa:79}.

The network length $L_{\m N}$ can be taken equal to $a$; it disappears in the
final result for $\Lambda L_{\m N}^2$, which is scale-independent. For a hexagonal
close packed lattice, $c=(8/3)^{1/2} a$ so that the expression $\Lambda L_{\m
  N}^2$ in Eq.(\ref{ShearModulusGranatoLucke}) takes the value
$2^{-1/2}$. This value is smaller if the lattice is less densely packed. For a
cubic lattice, a similar derivation gives the often quoted geometrical limit
$\Lambda L_{\m N}^2=3$, a value much larger than for a hexagonal lattice. This
result reflects the paucity of easy glide directions in the latter case.
\begin{figure}[b]
  \includegraphics[width=60mm]{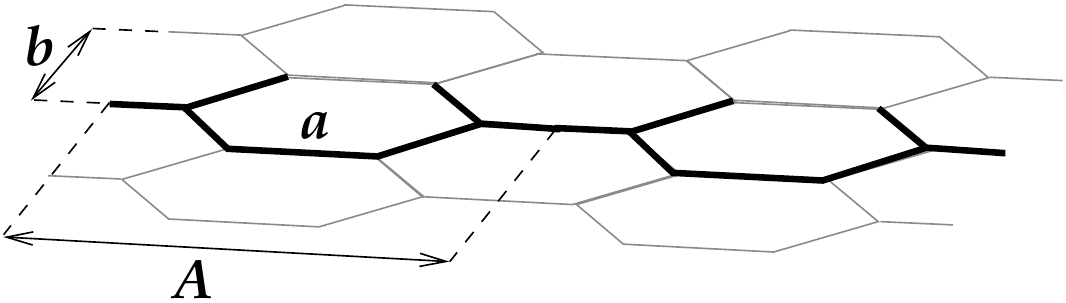}
  \caption{\label{hcp-dislocation} Dislocation network in a hexagonal
    structure. The network is formed of 2D hexagonal cells, of side length $a$
    in basal planes perpendicular to the $\hat{c}$-axis. The whole pattern can
    be generated from the elementary block in thick line of length $A$ by
    translations of moduli $A$ and $b$ as shown in the figure. }
\end{figure}

\end{document}